\documentclass[preprint,12pt,3p]{elsarticle}
\usepackage{epsfig}

\begin{document}

\newcommand{\be}{\begin{equation}}
\newcommand{\ee}{\end{equation}}
\newcommand{\ba}{\begin{eqnarray}}
\newcommand{\ea}{\end{eqnarray}}
\newcommand{\Gam}{\Gamma[\varphi]}
\newcommand{\Gamm}{\Gamma[\varphi,\Theta]}
\thispagestyle{empty}
\title{Comment on" Entanglement of two interacting bosons in a two-dimensional
isotropic harmonic trap"[ Physics Letters A 373 (2009) 3833-3837]\\
}
\author{Przemys\l aw Ko\'scik\\
Institute of Physics, Jan Kochanowski University\\
\'Swi\c{e}tokrzyska  15, 25-406 Kielce, Poland}

\begin{abstract}


The correct form of the Schmidt decomposition of the stationary wave functions for a system of two interacting particles trapped in a two-dimensional harmonic potential is given.

\end{abstract}

\maketitle

We note that  in Ref.\cite{int1}  the expansion of the two-particle wave function  (Eq.11) is mistakenly  interpreted   as the Schmidt decomposition.  The  mode functions appearing in it cannot be considered as the Schmidt orbitals  due to their incorrect normalization  in radial direction in two-dimensional (2D) space. It should be stressed that this mistake does not affect the validity of the results presented in  Ref. \cite{int1}  since  Eq.11  has not been used to their determination. However, because of the utility of the model in many areas of physics it is important to provide the correct Schmidt form decomposition that we derive below.

The system of two interacting particles trapped in a 2D isotropic harmonic potential,
irrespectively of the interaction potential between the particles, possess the stationary wave-functions $(m_{r}=M_{c}=0)$\footnote {The Hamiltonian of two interacting particles in the isotropic harmonic trap is separated
into center of mass (c.m.) and relative (rel) motion. The stationary states of this system may be chosen  as $\Psi_{n,m_{r},N,M_{c}}(\vec{\rho}_{1},\vec{\rho}_{2})=\psi^{rel.}_{n}(\rho)e^{im_{r} \varphi_{rel.}}\psi^{c.m.}_{N}(\varrho)e^{ iM_{c} \varphi_{c.m.}}$, where the functions $\psi^{rel}_{n}$ and $\psi^{c.m.}_{N}$  are solutions of  the radial Schr\"{o}dinger equations rel. and c.m., respectively.}  that  depend only on $\rho$ and $\varrho$, where
$|\vec{\varrho}|=|(\vec{\rho}_{1}+\vec{\rho}_{2})/2|=\varrho(\rho_{1},\rho_{2},cos(\varphi_{2}-\varphi_{1}))$ and
$|\vec{\rho}|=|\vec{\rho}_{2}-\vec{\rho}_{1}|=\rho(\rho_{1},\rho_{2},cos(\varphi_{2}-\varphi_{1}))$.
Being the function of $\rho$ and $\varrho$ only, $\Psi$ is  symmetric under permutation of the particles ($(\rho_{1},\varphi_{1})\longleftrightarrow(\rho_{2},\varphi_{2})$) and may be assumed to be real for simplicity.
An application the method of Ref. \cite{int2} to this case results in
     \be \Psi(\rho,\varrho)=\sum_{m=-\infty...\infty}{A_{m}(\rho_{1},\rho_{2})\over
     \sqrt{\rho_{1}\rho_{2}}}e^{im\varphi_{1}}e^{-im\varphi_{2}},
\label{ddd1}\ee
where to ensure correct normalization in radial directions we have introduced  $\sqrt{\rho_{1}\rho_{2}}$ which is the crucial difference  from Eq. 11 of  Ref. \cite{int1}.
The   function  $A_{m}(\rho_{1},\rho_{2})$ is given by
the following integral
\begin{eqnarray}
A_{m}(\rho_{1},\rho_{2})=\sqrt{\rho_{1}\rho_{2}}\int_{0}^{2\pi}\int_{0}^{2\pi}\Psi(\rho,\varrho)e^{im(\varphi_{2}-\varphi_{1})}d\varphi_{1}d\varphi_{2}=\nonumber\\
\sqrt{\rho_{1}\rho_{2}}\int_{0}^{2\pi}\int_{0}^{2\pi}\Psi(\rho,\varrho)cos(m(\varphi_{2}-\varphi_{1}))d\varphi_{1}d\varphi_{2}
\label{rrr},\end{eqnarray}
where the simplification has  been achieved by elementary symmetry considerations.
    Being real and symmetric,  the function $A_{m}(\rho_{1},\rho_{2})$ has the following Schmidt form
\be A_{m}(\rho_{1},\rho_{2})=\sum_{s=0}\kappa_{s,m}{\chi^{(m)}_{s}(\rho_{1})}{\chi^{(m)}_{s}(\rho_{2}) },\label{dds}\ee
where  the coefficients $\kappa_{s,m}$  and the orbitals $\chi^{(m)}_{s}(\rho)$   satisfy the integral
equation
$$ \int_{0}^{\infty} A_{m}(\rho_{1},\rho_{2})\chi^{(m)}_{s}(\rho_{2})d\rho_{2}=\kappa_{s,m}\chi_{s}^{(m)}(\rho_{1}).$$ The  family $\{\chi^{(m)}_{s}(\rho)\}_{s=0}$ forms a complete and orthogonal set
($\int_{0}^{\infty}\chi^{(m)}_{s}\chi^{(m)}_{s^{'}}d\rho=C\delta_{ss^{'}}$).
Using the expansion (\ref{dds}) the final form of the decomposition of the wave  function $\Psi$ now  reads
\be
\Psi(\rho,\varrho)=\sum_{m=-\infty...\infty\atop s=0}\kappa_{s,m}v_{s,m}(\rho_{1},\varphi_{1})v_{s,m}^{*}(\rho_{2},\varphi_{2}),
\label{ddd}\ee
where
\be v_{s,m}(\rho,\varphi)={\chi^{(m)}_{s}(\rho)\over \sqrt{\rho}}e^{im\varphi}\label{pp}.\ee
Since the orbitals (\ref{pp})  satisfy the condition of orthogonality in 2D space
$$ \int_{0}^{\infty}\int_{0}^{2\pi}\rho v_{s,m}^{*} v_{s^{'},m^{'}}d\rho d\varphi=2\pi
\delta_{m,m^{'}}\int_{0}^{\infty}\chi^{(m)}_{s}\chi^{(m)}_{s^{'}}d\rho=2\pi C\delta_{mm^{'}}\delta_{ss^{'}},$$
we recognize them as the Schmidt modes (natural orbitals). One can point out that they are the eigenfunctions of the angular momentum operator $\hat{L}_{z}$.

In conclusion,  we have presented in details a procedure of obtaining   the Schmidt decomposition  of the two-particle wave function that is a function
of distances $\rho$ and $\varrho$ only.

\end{document}